\documentclass[12pt,aps,prb,preprint,tightenlines]{revtex4} 
\usepackage{latexsym,amssymb,amsfonts,amsbsy,amsmath}
 \usepackage[pdftex]{graphicx}
\DeclareGraphicsExtensions{.pdf, .jpg}
\def\Hm{\ensuremath{\mathrm{H}^-}}
\def\vec#1{\boldsymbol{#1}}
\def\d{\mathrm{d}}
\begin{document}
\title{Two-electron atoms, ions and molecules}

\author{Hallstein H{\o}gaasen}
\email{hallstein.hogasen@fys.uio.no}
\affiliation{Department of Physics
University of Oslo\\
Box 1048  NO-0316 Oslo Norway}

\author{Jean-Marc Richard}
\email{jean-marc.richard@lpsc.in2p3.fr}
\affiliation{Laboratoire de Physique Subatomique et de Cosmologie,\\
 Universit{\'e} Joseph Fourier, CNRS-IN2P3, INPG\\
53, avenue des Martyrs, F-38036 Grenoble cedex, France}
\author{Paul Sorba}
\email{sorba@lapp.in2p3.fr}
\affiliation{Laboratoire d'Annecy-le-Vieux de Physique Th{\'e}orique,\\
UMR 5108, Universit\'e de Savoie, CNRS\\
9, chemin de Bellevue, B.P. 110, F-74941 Annecy-le-Vieux cedex, France}

\date{\today\\[25pt]}

\begin{abstract}
The quantum mechanics of two-electron systems is reviewed, starting with the ground state of the helium atom and helium-like ions, with central charge $Z\ge 2$. For $Z=1$, 
demonstrating the stability of the negative hydrogen ion, \Hm, cannot be achieved using a mere product of individual electron wave functions, and requires instead an explicit account for the anticorrelation among the two electrons. The wave function proposed by Chandrasekhar is revisited, where the permutation symmetry  is first broken and then restored by a counter-term.  More delicate problems can be studied using the same strategy: the stability of hydrogen-like ions $(M^+,m^-,m^-)$ for any value of the proton-to-electron mass ratio $M/m$; the energy of  the lowest spin-triplet state of helium and helium-like ions;  the stability of the doubly-excited hydrogen ion with unnatural parity. The positronium molecule $(e^+,e^+,e^-,e^-)$, which has been predicted years ago and discovered recently, can also be shown to be stable against spontaneous dissociation, though the calculation is a little more involved. Emphasis is put on symmetry breaking which can either spoil or improve the stability of systems.
\end{abstract}

\maketitle
%
\section{Introduction}\label{se:intro}
The chapter on two-electron atoms or ions is of great importance when teaching quantum mechanics, and usually the opportunity of a transition from simple binary systems to more complicated structures, with  examples of application of perturbation theory and variational methods.

Historically, understanding the two-electrons atoms was crucial to demonstrate that the theory of quanta was not just an ansatz that works fortuitously for the case of the hydrogen atom. Indeed, while the Bohr--Sommerfeld quantization method accounts efficiently for the one-electron atoms, it first faced serious difficulties for the description of helium. Then Heisenberg\cite{1926ZPhy...39..499H} and other pioneers (for refs., see, e.g., the book by Bethe and Salpeter\cite{1957qmot.book.....B}) showed that the helium atom can be well described in the framework of the new quantum mechanics.

However, binding two electrons to a helium nucleus or a heavier nucleus with charge $Z\ge 2$ is rather obvious, as the first attached electron leaves a positively-charged kernel that easily traps the  second electron. The problem here is  to calculate approximately the energy spectrum and the associated wave functions and  not to demonstrate the existence of bound states.
It is thus unfortunate that many textbooks, even among the best ones, are restricted to the case of helium and do not discuss the more challenging case of $Z=1$, i.e., the negative hydrogen ion. Noticeable exceptions are Refs.~\onlinecite{Griffiths:885094,Mandl:308811,Park:308810,Kroemer-MQ}.

The negative hydrogen ion, \Hm, enters a variety of physical, chemical, biological and geological processes.\cite{HM}\@ In astrophysics, it plays a role at Sun's surface, and its absorption and emission properties have been studied by Chandrasekhar in a series of papers.\cite{2001qpsw.book.....W}\@  In some laboratories, there are nowadays beams of \Hm. Intense beams of \Hm\ are foreseen for future nuclear-fusion devices.\cite{Fusion}\@  When teaching few-body quantum mechanics, it is the simplest prototype of fragile structure, at the edge between binding and non-binding, which cannot be described by simple tools such as Hartree wave-functions, however efficient are these methods to account for the properties of well-bound systems. Other examples are atomic clusters made of noble-gas atoms, or Borromean nuclei with two weakly-bound peripheral neutrons.

Intimately related to \Hm\ is the positronium molecule $(e^+,e^+,e^-,e^-)$ predicted in 1945 by Wheeler\cite{Wheeler:1945} and discovered only recently.\cite{2007Natur.449..195C}\@ Demonstrating its stability against dissociation into two positronium atoms can be done with a generalization of the Chandrasekhar wave function, though  the calculation becomes slightly more intricate. This molecule has many symmetries. It can be seen that breaking particle identity and breaking charge conjugation have dramatically different effects on its stability. In the former case, it quickly disappears, while it the later case, it is reinforced. In particular, the stability of the hydrogen molecule can be -- somewhat paradoxically -- demonstrated as a consequence of the  stability of the positronium molecule. This is of course at variance with the more physical starting point of two infinitely massive protons, but illustrates the importance of symmetry breaking which enters many other fields of physics.

This paper is aimed at reviewing what can be taught on the quantum mechanics of the two-electron atoms and molecules at the elementary or more advanced level. 
We begin in Sec.~\ref{two-at-gs} with the ground state of two-electron atoms and ions, which is a spin-singlet configuration.  The easiest case of a central charge $Z\ge 2$ is briefly reviewed, before discussing the case of the hydrogen ion with $Z=1$. We focus on the beautiful solution proposed by Chandrasekhar,\cite{1944ApJ...100..176C}  which is a mere product of single-electron wave functions with \emph{different} range parameters, supplemented by a counter-term in which the two electrons are interchanged, so that the overall permutation symmetry is restored. Two other levels of helium-like systems are presented in Sec.~\ref{two-at-ex},  the lowest spin-triplet state, whose orbital wave function is antisymmetric, and  the unnatural-parity state of the hydrogen ion which is very loosely bound below its threshold. In Sec.~\ref{two-mol},  the case of the positronium molecule is presented,  as well as some of  its less symmetric variants. After a brief summary is  in Sec.~\ref{se:concl}, some details about the calculation of the matrix elements are given in Appendix.
\section{The ground state of two-electron atoms and ions}\label{two-at-gs}
We consider first the non-relativistic Hamiltonian describing two electrons of mass $m$ and charge $e$ around a fixed charged $Ze$,
\begin{equation}\label{eq:H}
H =\frac{\vec p_1^2}{2m}+\frac{\vec p_2^2}{2m}-\frac{Ze^2}{r_1}-\frac{Ze^2}{r_2}+\frac{e^2}{r_{12}}~,
\end{equation}
with $r_{12}=|\vec r_2-\vec r_1|$.  
The Coulomb problem has very simple scaling properties: the  energies are proportional to $e^4 m/\hbar^2\simeq 27.211\;\mathrm{eV}$ and the distances to $\hbar^2/(m e^2)$. We shall give all results  in natural units which correspond to treating (\ref{eq:H}) as if  $m/\hbar^2=e^2=1$.  The
orbital wave function should be antisymmetric for a spin triplet,  and symmetric for a spin singlet, as the ground state we shall consider first.
\subsection{The helium atom and the heavier ions}
The case of $Z\ge2$ is treated in most textbooks. Hence we shall give only a minimal review, for the sake of completeness.
If the last term of (\ref{eq:H}) is omitted, the Hamiltonian is exactly solvable, and for the ground state, the unperturbed energy is $E_0=-Z^2$ 
and the wave function $\Psi_Z=(Z^3/\pi)\,\exp[-Z (r_1+r_2)]$.  To first order, the energy is approximated and upper bounded by $E_0+E_1=-Z^2+\langle\Psi_Z|r_{12}^{-1}|\Psi_Z\rangle$ (
$E_0+E_1$ is the variational energy corresponding to the trial wave function $\Psi_Z$).

The matrix element $\langle\Psi_Z | r_{12}^{-1} | \Psi_Z\rangle$ is routinely estimated by a partial-wave expansion. It is sufficient, as done, e.g., by Peebles,\cite{Peebles:239304}  to evoke the Gauss theorem, which states that the potential created at distance $r_2$ by a spherical shell $\delta q_1$ of radius $r_1$ is $\delta q_1/r_2$ if $r_1<r_2$ and $\delta q_1/r_1$ if $r_1>r_2$. This gives%
\begin{equation}\label{eq:E1}
E_1=4\,Z^6\,\int\limits_0^\infty \exp(-2 Z r)\,r^2\,\d r\left[\int\limits_0^r \frac{\exp(-2 Z r')}{r}\,r'^2\,\d r'+\int\limits_r^\infty \frac{\exp(-2 Z r')}{r'}\,r'^2\,\d r'\right]=\frac{5 Z}{8}~.
\end{equation}
Then for $Z=2$, one obtains an energy $-2.75$, to be compared to $E=-2.90372\ldots$ from the most sophisticated estimates,\cite{Lin95,2000PhRvA..61f4503K} and the lowest dissociation threshold $E_\text{th}=-2$. However, it is easily checked that this approach requires $Z> 5/4$ to bind two electrons.

An easy  and pedagogically instructive improvement  consists of replacing $\Psi_0$ by 
\begin{equation}\label{eq:eff-q}
\Psi_\alpha=(\alpha^3/\pi)\,\exp[-\alpha (r_1+r_2)]~,
\end{equation}
where $\alpha$ is a variational parameter, whose value measures the effective charge seen by each electron. The matrix elements are the same as for $\alpha=Z$, and the variational energy reads
\begin{equation}\label{eq:E2}
\widetilde{E}=\min_\alpha\left[\alpha^2-2Z\,\alpha+\frac{5\alpha}{8}\right]=-\left(Z-\frac{5}{16}\right)^2~,
\end{equation}
the minimum being reached for an effective charge $\alpha=Z-5/16$. 
For $Z=2$, this gives an improved $\widetilde E\simeq-2.8477$. Still binding with this wave function is demonstrated only for $Z\ge1.067$. Thus  $Z=1$ requires another treatment, as described in the following section.
\subsection{The negative hydrogen ion}
Variational wave functions that bind  \Hm\ have been written down by Bethe, Hylleraas and several others. See, e.g., the book by Bethe and Salpeter.\cite{1957qmot.book.....B}\@  For instance, a correlation factor $(1+ \beta r_{12})$ or $\exp(\gamma r_{12})$ can be inserted into the wave function~(\ref{eq:eff-q}). The most elegant solution is perhaps that of 
Chandrasekhar,\cite{1944ApJ...100..176C,2001qpsw.book.....W} which reads (unnormalized)
\begin{equation}\label{eq:chan}
\Phi=\exp(-a\, r_1-b\,r_2)+\epsilon\, \exp(-b\, r_1-a\,r_2)~,\quad \epsilon=+1~,
\end{equation}
where the permutation symmetry is explicitly broken by  $a\neq b$ and restored by the second term.  As compared to the standard shell-model wave function (\ref{eq:eff-q}), sometimes labelled $(1s)^2$, this wave-function is named  ``unrestricted'' by Goddard\cite{PhysRev.172.7} who gives a generalization.

The matrix elements of $\Phi$ involve the same basic integrals as for the simpler wave function $\Psi_\alpha$, and it is straightforward to derive, for the kinetic ($T$) and potential ($V$) energy and for the normalization ($N$),
\begin{equation}\label{eq:MECh}
\begin{aligned}
\overline{E}(a,b)&=\frac{\langle\Phi | H | \Phi \rangle}{\langle\Phi | \Phi \rangle} =\frac{T+V}{N}~,\\
N&=\frac{1}{8 a^3b^3}+\frac{8\,\epsilon}{(a+b)^6}~,\qquad\qquad
T=\frac{1}{16 a b^3}+\frac{1}{16 a^3 b}+\frac{8ab\,\epsilon}{(a+b)^6}~.\\
V&=-\frac{Z}{8a^2b^3}-\frac{Z}{8 a^3 b^2}-\frac{8Z\,\epsilon}{(a+b)^5}+\frac{5\,\epsilon}{2(a+b)^5}+\frac{a^2+3 ab+b^2}{8 a^2 b^2 (a+b)^3}~,
\end{aligned}
\end{equation}
where the attractive terms (proportional to $Z$) are supplemented by the contribution from $1/r_{12}$.

\begin{figure}[!thbc]
 \begin{minipage}{.65\textwidth}
 \centerline{\includegraphics[width=.85\textwidth]{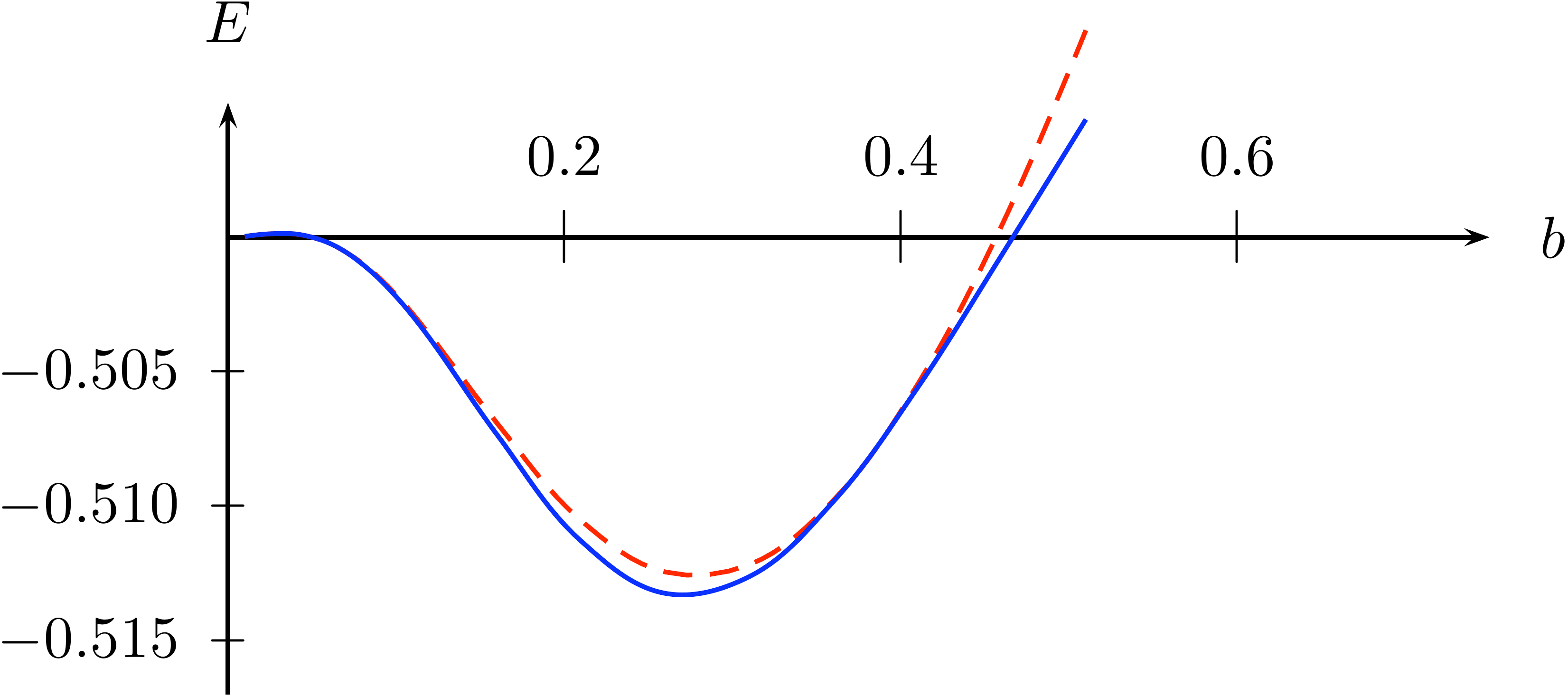}} \end{minipage}
 \hfill
 \begin{minipage}{.32\textwidth}
 \caption{\label{fig:chandra} One parameter minimization of the variational energy $\overline E(a,b)$ of \Hm, obtained from the Chandrasekhar wave function: with  the approximation of a frozen $a=1$ (dashed curve) and, without approximation but  using the virial theorem, which removes one parameter (solid curve).}
 \end{minipage}
 \end{figure}

As \Hm\ is weakly bound, the physical picture is that of one electron far away, and the other one  near the nucleus, feeling an almost unscreened Coulomb potential. This suggests the following approximation: one 
freezes out $a=Z=1$, i.e., assumes that one of the electrons is unperturbed, and  varies $b$, to get a first minimum $\overline{E}(1,b_0)\simeq -0.5126$ that already establishes binding! This minimum is reached for $b_0\simeq0.279$. See Fig.~\ref{fig:chandra}, dashed curve.

\begin{figure}[!bhtc]
 \begin{minipage}{.55\textwidth}
\includegraphics[width=.75\textwidth]{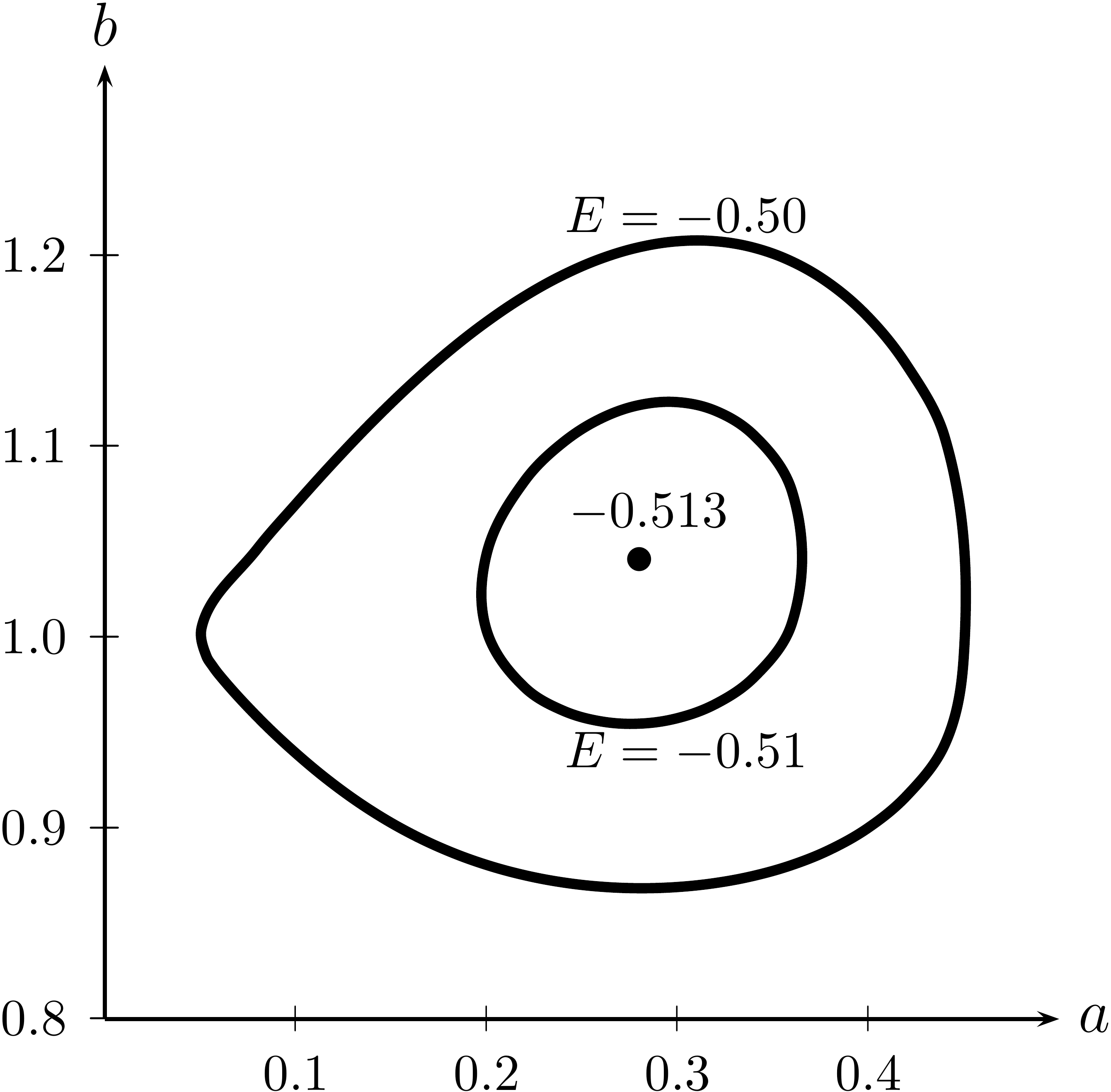}
\\
 \end{minipage}
 \hfill
 \begin{minipage}{.32\textwidth}
 \caption{\label{fig:contour} Contour plot of the variational energy $\overline{E}(a,b)$, given by Eq.~(\protect\ref{eq:MECh}), of the Chandrasekhar wave function (\ref{eq:chan}). The symmetric part where $a<b$ is not shown.}
 \end{minipage}
 \end{figure}

Using any standard minimization software easily leads to the best minimum
 $\overline{E}(a_1,b_1)\simeq -0.5133$ for $a_1\simeq1.039$  (indeed, very close to the previous approximate $a=1$) and $b_1\simeq 0.283$ (or $a_1\leftrightarrow b_1$). For comparison, the best non-relativistic energy for an infinitely massive proton gives\cite{PhysRev.112.1649,Lin95}  about $E=-0.52775$. As seen in Fig.~\ref{fig:contour}, the minimum is not extremely sharp, however, the stability criterion $\overline{E}(a,b)< 0.5$  clearly requires $b \ll a$ (or $a\ll b$).

The task of minimizing $\overline E(a,b)$ or any similar variational energy can be simplified by using the virial theorem, which also holds for the best 
 variational solution, with the mild restriction that the set of trial functions is globally invariant under rescaling. This was  noticed very early.\cite{1929ZPhy...54..347H,1930ZPhy...63..855F}\@  See, also, Refs.~\onlinecite{hoor:647,2005PhR...413....1A,liu:1202}. A simple derivation of the virial theorem consists, indeed, to impose that in a rescaling $\Psi(\vec r_1, \ldots)\to \lambda^{-3n/2} \,\Psi(\lambda\vec r_1,\ldots)$, where $n$ is the number of internal variables and the factor keeps the normalization,  the expectation value of the Hamiltonian remains stationary near $\lambda=1$. This obviously works  for both the exact solution or the best variational approximation in a given set. For instance, in the case of the Chandrasekhar wave function,  one can set $a=a_0(1+x)$, $b=a_0(1-x)$. For given $x$, the minimization over the overall scale $a_0$ fixes  the proper balance of kinetic and potential energy, as required by the virial theorem.  One is left with minimizing 
 \begin{equation}\label{eq:ener-vir}
-\frac{V^2}{4 NT}~,
 \end{equation}
 over the single variable $x$,  to recover the  minimum  at $E\simeq-0.5133$, as shown in Fig.~\ref{fig:chandra} (solid curve).

Some years ago, another trial wave function was proposed, in a paper\cite{srivastava:462} where the work of Chandrasekhar is not cited. It is the very compact 
 $\psi=\exp[-a r_{\scriptscriptstyle<}-b r_{\scriptscriptstyle>}]$, where $r_{\scriptscriptstyle<}=\min(r_1,r_2)$ and $r_{\scriptscriptstyle<}=\max(r_1,r_2)$. The calculation of the matrix elements of $\psi$ is similar to these of (\ref{eq:chan}). Optimizing $a$ and $b$  with or without the help of the virial theorem leads to a variational energy $-0.506$ which demonstrates the stability of \Hm, but gives less binding than the wave function (\ref{eq:chan}), due to the unphysical discontinuity of the radial derivatives at $r_1=r_2$.
 \subsection{Varying the proton charge}
 The method used  for \Hm\ can be applied to other values of the central charge $Z$.
 One can first return to $Z\ge 2$, where  the wave function (\ref{eq:chan}) gives an energy $\overline E\simeq-2.8757$, instead of $\widetilde E\simeq-2.8477$ from the factorized wave function (\ref{eq:eff-q}). As $Z$ increases, the improvement becomes less and less significant, i.e., the factorized wave function (\ref{eq:eff-q}) works almost equally well. See Tables~\ref{tab:triplet} and \ref{tab:or}.
 
  \begin{table}[!!!htbc]
 \caption{\label{tab:triplet}%
 Binding energies (in natural units) for a series of central charge $Z$ and electron spin $S$. The experimental energy $E_\text{exp}$ is taken from the current data bases \protect\cite{2002crc..book.....L,Data}, and compared to the best non-relativistic calculation with an infinitely massive nucleus, $E_\text{NR}$ \cite{PhysRev.112.1649,PhysRevE.62.8740,0953-4075-38-17-013}, the simplest Hartree--Fock type of calculation with an effective charge, $E_\text{fac}$  corresponding to (\ref{eq:eff-q}) for $S=0$ and (\ref{eq:shell-m}) for $S=1$, and the energy from the Chandrasekhar wave function (\ref{eq:chan}) with $\epsilon=+1$ for $S=0$ and $-1$ for $S=1$, with optimized range parameters $a$ and $b$.} 
 \begin{center}
 \begin{tabular}{cccccccc}
\hline\hline
 $Z$  & $S$ & $E_\text{exp}$ & $E_\text{NR}$ &  $E_\text{fac}$&    $E_C$    & $a$   & $b$ \\ 
\hline
1    &  0  &  $-0.5274$    & $-0.5277$  & $-0.4727$       &   $-0.5133$   & $1.04$   & $0.28 $   \\
2    &  0  &  $-2.9034$    & $-2.9037$  & $-2.8477$       &   $ -2.8757 $  &  $2.18$   & $1.19$       \\
2    &  1  &  $-2.1750$    & $-2.1752$  & $-2.1666$        &   $-2.1607$  &  $1.97$     & $0.32$     \\
3    &  0  &  $-7.2800$    & $-7.2799$  & $-7.2227$       &   $-7.2488$    &  $3.29$   & $2.08$     \\
3    &  1  &  $-5.1103$    & $-5.1107$  & $-5.1026$        &   $-5.0718$   &  $2.93$   & $0.60$     \\
4    &  0  &  $-13.657$    & $-13.656$  & $-13.598$       &   $-13.623$   & $4.39$     & $2.98$     \\
4    &  1  &  $-9.2988$    & $-9.2972$  & $-9.2892$        &   $-9.2240$   &  $3.89$    & $0.88$     \\
8    &  0  &  $-59.195$    & $-59.157$  & $-59.098$      &   $-59.122$   &   $8.68$   & $6.69$     \\
8    &  1  &  $-38.579$    & $-38.545$   & $-38.537$     &   $-38.233$   &  $7.73$     & $2.00$     \\
\hline\hline
\end{tabular}
\end{center}
\end{table}
 
 Alternatively, one can investigate how far one can decrease $Z$ without breaking the system.
It can be checked that stability remains down to $Z\simeq 0.949$ with the wave function (\ref{eq:chan}).  In the literature, the most sophisticated estimate of the critical  charge to bind two electrons is $Z_c\simeq 0.9107$, from the so-called $1/Z$ expansion. By scaling, the Hamiltonian is rewritten as 
\begin{equation}\label{eq:Hsca}
\frac{H}{Z^2} =\frac{\vec p_1^2}{2 m}+\frac{\vec p_2^2}{2 m}-\frac{1}{r_1}-\frac{1}{r_2}+\frac{1}{Z\,r_{12}}~,
\end{equation}
and the energy is expanded in powers of $1/Z$. From (\ref{eq:E1}), the first terms are
$Z^2[-1+5/(8Z)+\cdots]$. Elaborate studies\cite{PhysRevA.41.1247,PhysRevA.51.1080,PhysRevA.52.1942}  (beyond the scope of the present article)  have shown that
\begin{itemize}  
\item
The expansion looks like
\begin{equation}\label{eq:Z1}
E=-Z^2\biggl(1-{5\over8Z}
+{0.157666429\over Z^2}  -{0.008699032\over Z^3}+{0.000888707\over
Z^4}+\cdots\biggr)~.
\end{equation}
\item The convergence of the series is associated to well isolated bound states lying below the threshold. When the convergence breaks, the binding is lost. The radius of convergence $1/Z\simeq 1.098$ leads to the critical charge $Z_c\simeq 0.9107$.
\end{itemize}
 \subsection{Proton recoil}
 Coming back to unit charges, one might examine the effect of  a finite mass for the proton, and more generally, study the stability of $(M^\pm,m^\mp, m^\mp)$ as a function of the mass ratio $M/m$. The Hamiltonian now reads
\begin{equation}\label{eq:HMm}
H =\frac{\vec p_1^2}{2m}+\frac{\vec p_2^2}{2m}+\frac{\vec p_3^2}{2M}-\frac{1}{r_1}-\frac{1}{r_2}+\frac{1}{r_{12}}~,
\end{equation}
 The negative hydrogen ion corresponds to $M/m\to \infty$, the positronium ion, $\mathrm{Ps}^-$, to $M/m=1$, and the molecular-hydrogen ion, $\mathrm{H}_2{}^+$, to $M/m\to 0$.

Hill\cite{hill:2316} has shown that  Chandrasekhar's wave function (\ref{eq:chan}) demonstrates binding for any $M/m$. When $M/m$ is varied, the minimum is reached with the same $b/a$ and the same quality of binding, as measured by the ratio of the best variational energy to the threshold energy.

The proof is rather straightforward. One introduces standard coordinates
\begin{equation}\label{eq:new-coo}
\vec{x}=\vec r_1-\vec r_3~,\quad \vec{y}=\vec r_2-\vec r_3~,
\quad \vec R=\frac{m \vec r_1+m \vec r_2
+M \vec r_3}{2 m+M}~,
\end{equation}
and the conjugate momenta
\begin{equation}\label{eq:conj-mom}
\vec p_x=\frac{(m+M)\vec p_1-m(\vec p_3+\vec p_2)}{M+2m}~,\quad
\vec p_y=\frac{(m+M)\vec p_2-m(\vec p_3+\vec p_1)}{M+2m}~,\quad
\vec P=\vec p_1 +\vec p_2+\vec p_3~,
\end{equation}
 in terms of which the Hamiltonian becomes
 \begin{equation}
 H=\frac{\vec P^2}{4 m+ 2 M}+\left[ \frac{\vec p_x^2}{2\mu}-\frac{1}{|\vec x|}+
 \frac{\vec p_y^2}{2\mu}-\frac{1}{|\vec y|}+ \frac{1}{|\vec{x}-\vec{y}|}\right]+\frac{1}{M}\, \vec p_x.\vec p_y
 \end{equation}
 Once the center-of-mass motion removed, one is left with the Hughes--Eckart term, and, in the bracket, a rescaled version of \Hm\ with an infinitely massive proton and two electrons whose mass is decreased from  $m$ to $\mu=m M/(m+M)$.  This Hamiltonian in the bracket, if alone, gives stability with respect to the decay into a $(M^+,m^-)$ atom and an isolated negative charge. However, the Hughes--Eckart term has zero expectation value within the wave function (\ref{eq:chan}) with $r_1\to |\vec{x}|$ and $r_2\to |\vec y|$, or in any similar wave function in which there is no dependence upon the angle between $\vec{x}$ and 
 $\vec{y}$. Hence the Chandrasekhar wave function gives the same energy as for the original \Hm, apart from an overall scaling factor $\mu/m$.
 \subsection{Symmetry breaking in three-charge systems}
So far, we studied configurations of the type $(M^+,m^-,m^-)$ with two identical negative particles. One might address the question of stability of more general mass configurations 
$(M^+,\linebreak[2]{m_1^-},m_2^-)$. The most general case is discussed in the literature,\cite{2005PhR...413....1A} with stable configurations such as Ps$^-$ or \Hm, and unstable ones such as $(p,\bar p,e^-)$. We shall restrict the discussion here to small differences between $m_1$ and $m_2$.

It is known that symmetry breaking lowers  the  ground state. For instance, in one-dimensional quantum mechanics, $h=p^2+x^2+\lambda x$ has a ground state at $\epsilon =1-\lambda^2/4$ shifted down by the odd term. More generally, if $H=H_0+\lambda H_1$, with $H_0$ even and $H_1$ odd under some symmetry, then the variational principle applied to $H$ with the even ground state $\Psi_0$ of $H_0$   as trial wave function, gives for the ground state $E(\lambda)\le E(0)$ provided that $\langle \Psi_0| H_1| \Psi_0\rangle=0$.

Hence, if the $(M^+,m_1^-,m_2^-)$ Hamiltonian is split into\cite{PhysRevA.71.024502}
\begin{equation}
H(M^+,m_1^-,m_2^-)=H(M^+,\mu^-,\mu^-)+ \frac{m_1^{-1}-m_2^{-1}}{4}\left(\vec p_1^2-\vec p_2^2\right)~,
\end{equation}
where $\mu$ is the average inverse mass, the ground-state energy is shifted down by the second term, i.e., $E(M^+,m_1^-,m_2^-)\le E(M^+,\mu^-,\mu^-)$. But the gain is only at second order in $m_1^{-1}-m_2^{-1}$, and meanwhile, the lowest threshold decreases at first order, with $E_2(M,m_2)< E_2(M,\mu)$ if $m_1<m_2$ and thus $\mu<m_2$. Not surprisingly, the net result is that stability deteriorates as the two negative charges are given different masses. 

It is an interesting exercise to adapt the wave-function (\ref{eq:chan}) and to study the domain of stability of $(M^+,m_1^-,m_2^-)$ as a function of $m_1$ and $m_2$,  in the limit $M\to\infty$.
\section{First excitations of two-electron atoms and ions}\label{two-at-ex}
 \subsection{Spin-triplet ground state}
 If the wave function (\ref{eq:chan}) is used with $\epsilon=-1$, i.e., in its antisymmetric version, it becomes a trial wave function for the lowest spin-triplet state. For \Hm, this level is unstable. 
 Some results are shown in Table~\ref{tab:triplet} for $Z\ge 2$. In particular, one gets $E\simeq-2.16064$ for the lowest spin-triplet state of helium, to be compared to $E\simeq-2.17523$ from wave functions with many parameters. Also shown in this Table is the result obtained from the simplest alternative wave function that comes to mind, 
\begin{equation}\label{eq:shell-m}
\begin{aligned}
\Psi_{a,b}(r_1,r_2)&=\frac{\phi_{1s}(a, r_1)\phi_{2s}(b,r_2)-\phi_{2s}(b, r_1)\phi_{1s}(a,r_2)}{\sqrt{2}}~,\\
\phi_{1s}(a,r)&=\frac{a^{3/2}}{\sqrt{\pi}} \exp(-a r)~,\qquad \phi_{2s}(a,r)=\frac{a^{3/2}}{\sqrt{8\pi}} \,(1-a r/2)\,\exp(-a r/2)~.
\end{aligned}
\end{equation}
%
\def\phn{\phantom{1}} 
\begin{table}[hbtc]
\renewcommand{\arraystretch}{1.2}
\caption{Binding energies of H$^-$ ($Z=1$) and He ($Z=2$)
with an infinitely massive nucleus, obtained from the variational wave
function (\protect\ref{eq:changg}). For He, we also show the two
first excitations in the scalar sector: He$^*$(para) with the same spin
$S=0$ as the ground state, and He$^*$(ortho) with a symmetric $S=1$ and
thus an antisymmetric space wave function.} 
\label{tab:or}
\begin{center}
\begin{tabular}{cccccc}
 \hline\hline
 $N$ & $a_i,b_i,c_i$ & H$^-$ & He&He$^*$(para)&He$^*$(ortho)\\ 
\hline
1 & $a=b=Z$, $c=0$ &--0.375\phn\phn & --2.75\phn\phn&& \\
1 & $a=b$ , $c=0$ & --0.47266 & --2.84766 &&\\
1 & $a=b$, $c\ne0$ & --0.50790 & --2.88962&& \\
1 & $a\ne b$, $c=0$ &--0.51330 & --2.87566&&--2.16064\\
1 & $a\ne b$, $c\ne0$ &--0.52387 & --2.89953&&--2.16153\\
2 & $a\ne b$, $c\ne0$ &--0.52496 & --2.90185 &--2.14461&--2.17512\\
3 & $a\ne b$, $c\ne0$ &--0.52767 & --2.90328&--2.14538&--2.17521\\
4 & $a\ne b$, $c\ne0$ &--0.52771 & --2.90347&--2.14551&--2.17522\\
\hline
\multicolumn{2}{c}{``Exact''\protect\cite{Lin95}}& --0.52775 
&--2.90372&--2.14597&--2.17523\\
\hline\hline
\end{tabular}
\end{center}
\end{table}

For $a=b$, it is a standard normalized shell-model wave function, and 
corresponds the fifth column of Table \ref{tab:triplet}: the results are slightly better than these  the Chandrasekhar wave function.
If the above ``$(1s)(2s)$'' wave function is used with different range parameters $a$ and $b$ (and thus with non-orthogonal individual wave functions), corresponding to a  ``unrestricted'' Hartree--Fock wave-function in the notation of Goddard,\cite{PhysRev.172.7} it gives slightly better results, especially for small $Z$. In the case of large $Z$, the wave function (\ref{eq:shell-m}), with $a\to b\to Z$ becomes exact.
 \subsection{Towards a more accurate calculation}
For both the spin-singlet and the spin-triplet cases, the Chandrasekhar wave function, however astute, cannot describe completely the three-body ground-states such as \Hm. It can be improved by superposing more terms of the same kind.
For instance, Goddard\cite{PhysRev.172.7} considered a symmetrized combination of  products of $1s$, $2s$, \dots, $5s$ orbitals with different range parameters. For the ground-state with spin singlet, he got $E\simeq -0.5138$, which is a modest improvement as compared to  Chandrasekhar's result $E\simeq-0.5133$, which corresponds to restricting oneself to two $1s$ orbitals.

The most general scalar wave function depends on three variables, which can be chosen as the relative distances $r_1=r_{31}=y$, $r_2=r_{23}=x$ and $r_{12}=z$. Hence the  wave functions without explicit $r_{12}$ dependence will never approach the exact solution with arbitrary accuracy. 

Starting from  (\ref{eq:chan}), a natural extension is first 
\begin{equation}\label{eq:changg}
\Psi=\exp(-a \,x-b \,y- c \,z)\pm \{ a \leftrightarrow b \}~,
\end{equation}
and next, a superposition of such terms.  With a single term $(N=1)$,  one gets the results listed in Table~\ref{tab:or}.  Also shown are the improvement brought by superposing $N=2,\,3$ or $4$ such terms. For larger $N$, the numerical minimization becomes delicate, and requires clever tools, such as stochastic search of the parameters. Frolov, \cite{PhysRevA.58.4479} 
Korobov,\cite{2000PhRvA..61f4503K} and others have developed a systematics of expansions based on such exponential terms and have obtained extremely accurate results.

\subsection{Hydrogen ion with unnatural parity}\label{se:un}
 Another challenging problem deals with states of unnatural parity. 
 Consider again the $(p,e^-,e^-)$ system, though similar considerations could be developed in the four-body case. If one neglects the spins and intrinsic parities, the ground state has angular momentum and parity $0^+$. It is the only level of \Hm\ below the lowest threshold $\mathrm{H}(1s)+e^-$, as shown by Hill.\cite{hill:2316}

However, the state with quantum number $1^+$, i.e., unnatural parity, cannot decay into $\mathrm{H}(1s)+e^-$, at least as long as radiative corrections and spin-dependent effects are neglected. Its lowest threshold is $\mathrm{H}(2p)+e^-$ at $E_{\rm th}=-0.125$ in natural units. It has been discovered\cite{Hmoneplus} that the lowest state of \Hm\ with $1^+$ actually lies below this threshold, and the other calculations\cite{PhysRev.138.A1010,PhysRevLett.24.126,1979JChPh..71.4611J} have confirmed an energy $E\simeq-0.1253$. The question is to find the most economical way of demonstrating this binding. 

The simplest wave function bearing the right quantum numbers for this state is ($i$ is any projection of the vector product)
\begin{equation}\label{eq:psi-un}
\Psi(a,b,c)=(\vec{x}\times\vec{y})_i \left[\exp(-a x - b y - c z)+ \{a\leftrightarrow b\}\right]~.
\end{equation}
After angular integration, one is left with integrating a polynomial in $x$, $y$ and $z$ times an exponential, and the result can be deduced from a single generating function, as outlined in Appendix.  It can be checked, after optimization of the range parameters $a$, $b$ and $c$ (or two of them if one uses the virial theorem) that this wave function just fails to bind the unnatural state of \Hm. One needs  a superposition, say
\begin{equation}\label{eq:psi-un1}
\sum_i \alpha_i \,\Psi(a_i,b_i,c_i)~,
\end{equation}
For a given set of range parameters $\{a_i,b_i,c_i\}$, the coefficients $\alpha_i$, and the resulting energy are given by a generalized eigenvalue problem. Then the range parameters can be adjusted by standard techniques, if the number of terms is limited. If this number increases, special care is required, to avoid numerical instabilities. To simplify the minimization, one can extract the range parameters $a_i$, $b_i$ and $c_i$ from an arithmetic series $\alpha, \alpha+\beta, \alpha+2\beta, \ldots$, allowing the possibility of equal values, e.g., $b_i=c_i$. The minimization thus runs only on $\alpha$ and $\beta$. If $\alpha<0$ and $\beta >0$, which helps introducing some anticorrelation among the two electrons, one should impose $a_i+b_i>0$, $b_i+c_i>0$ and $c_i+a_i>0$. In the case of \Hm\ with $1^+$, we demonstrated the stability with a few terms and thus confirmed the earlier results.\cite{Hmoneplus,PhysRev.138.A1010,PhysRevLett.24.126,1979JChPh..71.4611J}

If one repeats the calculation in the case of Ps$^-$, one never reaches a variational energy below the $\mathrm{Ps}(2p)+e^-$ threshold. This confirms the conclusion by Mills, who found this state unbound.\cite{PhysRevA.24.3242}
\section{Two-electron molecules}\label{two-mol}
\subsection{The positronium molecule}
 In 1945, Wheeler suggested a variety of new states containing positrons, which could be stable in the limit where internal annihilation ($e^++e^-\to \gamma$'s) is neglected.\cite{Wheeler:1945}\@ Among the predictions was the positronium molecule, Ps$_2$, $(e^+,e^+,e^-,e^-)$. In 1946, Ore, then at Yale, tried very hard to calculate  this molecule, and concluded that  it is likely to be unstable.\cite{PhysRev.70.90}\@ However, the next year, Hylleraas and the very same Ore presented a beautiful proof of the stability,\cite{PhysRev.71.493} based on the wave function 
 \begin{equation}\label{eq:HO}
\Psi=\exp(-a r_{13}- b r_{14} -a r_{24} -b r_{23})+ \{a \leftrightarrow b\}~,
\end{equation}
which is an obvious generalization of  (\ref{eq:chan}). All the matrix elements can be calculated analytically.\cite{PhysRev.71.493,2005PhR...413....1A}\@  Some hints are given in Appendix.
With $a+b=1$ and $a-b=\beta$, the normalization, kinetic and potential energy read
\begin{eqnarray}
\label{4u:eq:Hylleraas-ev}
&& n={33\over16}+{33-22\beta^2+5\beta^4\over16(1-\beta^2)^3}~,\qquad
t={21\over8}-{3\beta^2\over2}
  +{21-6\beta^2+\beta^4\over8(1-\beta^2)^3}~,\\
&& v={19\over6}+{21-18\beta^2+5\beta^4\over4(1-\beta^2)^3}
 -{1\over(1-\beta^2)^2}\left[1-{5\beta^2\over8}-{1\over4\beta^4}
   +{7\over8\beta^2}+{(1-\beta^2)^4\over4\beta^6}\ln
{1\over1-\beta^2}\right]~.\nonumber
\end{eqnarray}
and using the virial theorem, $E=-v^2/(4 t n)$ should be minimized by varying $\beta$.
Though it does not include explicit dependence upon $r_{12}$ nor $r_{34}$, the wave function  (\ref{eq:HO}) suffices to establish binding at $E\simeq-0.5042$ below the threshold for spontaneous dissociation into two positronium atoms, at $E_{\rm th}=-0.5$. 
This energy has been much lowered by more and more sophisticated computations, \cite{PhysRevLett.80.1876}  to reach about $-0.51600$.
It was later realized that there are excited states, whose threshold is higher than two positronium atoms in the ground state, due to selection rules; for refs., see, e.g., Ref.~\onlinecite{puchalski:183001}.  
An indirect experimental evidence for the Ps$_2$ molecule was reported recently,\cite{2007Natur.449..195C}  62 years after  its  prediction! 
\subsection{Other molecules}
Once the positronium molecule is shown to be stable, one might study what happens for other mass configurations. Though the hydrogen molecule $(M^+,M^+,m^-,m^-)$ is better described from the large $M/m$ limit, i.e., the Born--Oppenheimer approximation, it is amazing that its stability can be understood  from the  $M=m$ limit. It is also rather instructive to study whether or not symmetry breaking does improve binding.

Indeed, a system $(\mu^+,\mu^-,\mu^-,\mu^-)$, i.e., any rescaled version of  Ps$_2$, has many symmetries: exchange of the positive or the negative particles, and overall charge conjugation. 
%

Consider first a breaking of permutation symmetry, for simplicity, identically in the positive and the negative sectors, keeping the average inverse mass $\mu^{-1}$ constant. This corresponds to writing the Hamiltonian as\cite{2005PhR...413....1A}
\begin{equation}
H(M^+,m^+,M^-,m^-)=H(\mu^+,\mu^+,\mu^-,\mu^-)+\frac{M^{-1}-m^{-1}}{4}(\vec p_1^2-\vec p_2^2
+\vec p_3^2-\vec p_4^2)~.
\end{equation}
The second term decreases the energy of the molecule. However, the same effect is observed as for the three-body ion:  the threshold decreases more substantially, benefitting from the property of two-body energies
\begin{equation}
E_2(M^+,M^-)+E_2(m^+,m^-)\le 2 E_2(\mu^+,\mu^-)~.
\end{equation}
Detailed numerical studies have shown that stability is lost for $M/m\gtrsim 2.2$ (or $m/M\gtrsim 2.2$)\cite{PhysRevA.55.200}.

If, instead,  charge conjugation is broken, i.e., if the mass configuration becomes $(M^+,M^+,\linebreak[1]{m^-,}m^-)$, the decomposition reads
\begin{equation}
H(M^+,M^+,m^-,m^-)=H(\mu^+,\mu^+,\mu^-,\mu^-)+\frac{M^{-1}-m^{-1}}{4}(\vec p_1^2+\vec p_2^2
-\vec p_3^2-\vec p_4^2)~,
\end{equation}
and again the four-body ground-state energy is lowered by the second term. Now, the threshold remains constant, at 
\begin{equation}
2 E_2(M^+,m^-)= 2 E_2(\mu^+,\mu^-)~,
\end{equation}
and thus the stability is improved. Indeed, the hydrogen molecule is bound by about 17\%
below the atom--atom threshold, whilst this is only about 3\% for 
the positronium molecule.
\section{Summary}\label{se:concl}
We briefly reviewed how the stability of the ground state of \Hm, and the lowest spin-triplet state of helium can be reached with rather simple wave functions, whose matrix elements can be estimated by straightforward calculus. A more delicate--and less advertised--problem is that of the stability of the unnatural parity states, which forces one to push further the variational expansion, in order to demonstrate binding.

The main message is that the Hartree--Fock method, i.e., a factorized wave functions with suitable (anti-) symmetrization is extremely efficient for deeply-bound systems, but fails for demonstrating the binding of states at the edge between stability and spontaneous dissociation.  This is also observed in nuclear physics: halo nuclei with a weakly bound external neutron and the Borromean
 nuclei with two external neutrons, require a dedicated treatment (A  Borromean 3-body system is bound while its subsystem are unstable. For instance, ${}^5\mathrm{He}=(\alpha,n)$ and $(n,n)$ are not bound, but ${}^6\mathrm{He}=(\alpha,n,n)$ is stable against any dissociation and only rely on $\beta$ decay to disintegrate.).

The strategy initiated by Hylleraas, Chandrasekhar and others consists of using a basis of functions where each term breaks permutation symmetry. The proper boson or fermion statistics is restored by superposing terms deduced by permutation. 
The same strategy guided Hylleraas and Ore when they derived the first proof of stability of the positronium molecule, and lies also underneath the most recent calculations of this system. 
For example, 
in their study of the positronium molecule,\cite{PhysRevLett.80.1876} Varga and Suzuki used a basis of correlated Gaussians, and their own variant of the parameter search.\cite{Suzuki98}\@
This method is more and more widely used in  quantum chemistry and other few-body problems. It consists, if $\vec x_1, \ldots\vec x_n$ are the internal variables, in describing the wave functions  as superpositions of states such as 
 \begin{equation}\label{eq:CG}
 \psi= \exp\left[-\hbox{$\sum_{i<j}a_{ij}\vec x_i.\vec x_j$} \right]+\cdots~,
 \end{equation}
where the ellipses are meant for terms deduced by permutation, charge conjugation or any other relevant symmetry which can be explicitly enforced.
The two-electron  atoms seems the best introduction to the advanced \textsl{ab-initio} calculations.

\appendix
\section{Calculation of the matrix elements}
In this appendix, we give some hints to estimate the matrix elements. These of the  wave functions (\ref{eq:eff-q}) and (\ref{eq:chan}) have been given explicitly.  Consider now its generalization $\phi=\exp(-a x - b y - cz)$, or say, $|a,b,c\rangle$,  whose symmetrized or antisymmetrized version  (\ref{eq:changg}) is used for spin-singlet and spin-triplet states, respectively. Here $\vec x=\vec r_2-\vec r_3$, $x=\|\vec x\|$, etc. 
The matrix elements are integrals over $xyz\,\d x\,\d y\, \d z$, restricted by the triangular inequality, and can be all deduced from the generic function
\begin{equation}\label{eq:generic}
F_3(\alpha, \beta,\gamma)=\iiint\limits_{|x-y|\le z \le x+y} \exp(-\alpha\, x -\beta\, y - \gamma\, z)\, \d x\,\d y\,\d z=\frac{4}{(\alpha+\beta)(\beta+\gamma)(\gamma+\alpha)}~,
\end{equation}
and its derivatives
\begin{equation}\label{eq:generic1}
G(i,j,k;\alpha,\beta,\gamma)=(-1)^{i+j+k}\frac{\partial^{\,i+j+k} F_3(\alpha, \beta,\gamma)}%
{\partial\alpha^i\, \partial\beta^j\,\partial\gamma^k}~,
\end{equation}
For instance, the normalization of (\ref{eq:chan}), besides a factor $8\pi^2$ due to trivial angular variables, reads
\begin{equation}\label{eq:generic2}
\langle a,b,c | a,b,c\rangle = G(1,1,1,2a, 2b, 2c)~,
\end{equation}
and any potential term is similar, e.g., 
\begin{equation}\label{eq:generic3}
\langle a,b,c |r_{12}^{-1}| a,b,c\rangle = G(1,1,0,2a, 2b, 2c)~,
\end{equation}
and this is easily extended  to non diagonal terms, with $2a\to a+a'$ etc.

Consider now the term $\vec p_1$ of the kinetic energy. It is a linear combination of gradients with respect to the distances,
\begin{equation}\label{eq:generic4}
\vec p_1=(-i) [\vec\nabla_z -\vec\nabla_x]~,
\end{equation}
this giving additional constant factors  and $\hat{\vec y}.\hat{\vec z}=(x^2-y^2-z^2)/(2 y z)$, namely,
\begin{multline}
\langle a,b,c |\vec p_1^2| a,b,c\rangle = (bb'+cc')\langle a,b,c | a,b,c\rangle\\
-\frac{bb'+cc'}{2}\left[G(3,0,0,\bar a,\bar b,\bar c)-G(1,2,0,\bar a,\bar b,\bar c)-G(1,0,2,\bar a,\bar b,\bar c)\right]~,
\end{multline}
where $2\bar a=a+a'$, etc.

For the wave function (\ref{eq:psi-un}), some angular integrals should be done beforehand, and one is left with similar integrals over $x$, $y$ and $z$.

We now consider the four-body problem, with a wave function of the type
\begin{equation}\label{eq:psi4}
\Psi=\exp(-a r_{13}- b r_{14}- c r_{23}-d r_{24})~.
\end{equation}
%
If $\vec r_{12}$, $\vec r_{13}$ and $\vec r_{23}$ are chosen as the internal coordinates, for a scalar wave function and a scalar operator that do not depend explicitly on $r_{34}$, 
one can work independently in he triangles $(1,2,3)$ and $(1,2,4)$ as done previously for the three-body systems, and after summing  over the trivial angular variables, the integrals run over
\begin{equation}
    {\rm d}\tau=r_{13}r_{14}r_{23}r_{24}\,{\rm d}r_{12}{\rm d}r_{13}{\rm d}r_{14}
                               {\rm d}r_{23}{\rm d}r_{24},
    \label{eq:integration-4body}
\end{equation}
A basic integral is 
\begin{equation}\label{eq:basicI-4}
\begin{aligned}
  F_4(a,b,c,d,u)&= \int\frac{{\rm d}r_{12}{\rm d}r_{13}{\rm d}r_{14}
                               {\rm d}r_{23}{\rm d}r_{24}}{r_{12}}
       \exp(-a r_{13}-b r_{23}-cr_{14}-dr_{24}-ur_{12})\\
      &= \frac{16}{ (a-b)(a+b)(c-d)(c+d)} \log\left[\frac{(b+c+u)(a+d+u)}{(a+c+u)(b+d+u)}\right]~.
        \end{aligned}
\end{equation}
where the triangular inequalities are more easily accounted for by using the variables 
$s_i=r_{1i}+r_{2i}$ and $t_i=r_{1i}-r_{2i}$ for $i=3,4$.
All matrix elements are related to $F$ and its derivatives. For instance, the normalization,  first attractive term, internuclear and electronic repulsion   of (\ref{eq:psi4}) are
\begin{equation}\label{eq:4-matel}
\begin{aligned}
 n(a,b,c,d)&= -\left. \frac{\partial F_4(a,b,c,d,u)}
 {\partial u \partial a \partial b \partial c \partial d }\right|_{u=0}~,
 \qquad
& v_{13}(a,b,c,d)&= \left.  \frac{\partial^4 
   F_4(a,b,c,d,u)}{ \partial u \partial b \partial c \partial d}\right|_{u=0}~, \\
v_{12}(a,b,c,d)&=\frac{\partial^4 
   F_4(a,b,c,d,0)}{ \partial a \partial b \partial c \partial d}~,
&v_{34}(a,b,c,d)&=v_{12}(a,c,b,d)~,
    \end{aligned}
\end{equation}
while for the kinetic energy of, e.g., the third particle, one gets
\begin{equation}
   \langle\Psi\vert\vec{p}_{3}^{2}\vert\Psi\rangle=
   (a^{2}+b^{2})\langle\varphi\vert\Psi\rangle-2 ab
   \langle\varphi\vert(r_{12}^{2}-r_{13}^{2}-r_{23}^{2})/(2 
   r_{13}r_{23})\vert\Psi\rangle~,
    \label{eq:p3sq}
\end{equation}
which can be expressed as a combination of derivatives of $F_4$. For a non-diagonal matrix elements between (\ref{eq:psi4}) and an analogous function with range parameters $a',\ldots d'$, 
the coefficients in the above expression become $a\,a'+b\,b'$ and $ a\,b'+a'\,b$, respectively, and the arguments of $F_4$ are taken to be $(a+a')/2$, \dots, $(d+d')/2$.

\end{document}